\documentclass{svjour2}
\newcommand{\be}{\begin{equation}}
\newcommand{\ee}{\end{equation}}
\usepackage{mathptmx}
\usepackage{latexsym}
\journalname{Journal of Statistical Physics}
\begin{document}

\title{In memory of Leo P. Kadanoff}
\author{Franz J. Wegner}
\institute{Institute for Theoretical Physics,
Ruprecht-Karls-University, Heidelberg}
\date{Received: Accepted:}
\maketitle

\begin{abstract}
Leo Kadanoff has worked in many fields of statistical mechanics.
His contributions had an enormous impact.
This holds in particular for critical phenomena,
where he explained Widom's homogeneity laws by means
of block-spin transformations and laid the basis for
Wilson's renormalization group equation.
I had the pleasure to work in his group for one year.
A short historically account is given.
\keywords{Renormalization group \and Block-spin transformation \and 
Critical phenomena \and Ising Model \and Duality}
\end{abstract}

\section{Introduction}

Leo Kadanoff has worked in many fields of Statistical Mechanics.

He started out in working on superconductivity in a thesis under the supervision of Paul Martin at Harvard.
Leo and Gordon Baym developed self-consistent approximations which 
preserved the conservation laws\cite{BayKad61}.
They published the widely used book {\it Conservation laws and 
correlation functions}\cite{KadBay62}.
Gordon Baym has reviewed his time with Leo in his contribution 
{\it Conservation laws and the quantum theory
of transport: The early days}.\cite{Baym}

After a number of papers related to superconductivity and transport
phenomena he became
interested in critical phenomena, where he contributed
essentially. Section 2 reviews shortly the situation in this field in 
the fifties and sixties of the last century.
Section 3 reviews the contributions
of Ben Widom and Leo Kadanoff in 1965 and 1966, which were two
important steps in the understanding of this field. This led to a
strongly growing interest in this field (sect. 4).
Finally, in 1971 Ken Wilson
developed the tool to calculate explicitly the critical behavior.
In the same year Rodney Baxter solved the eight-vertex model
and enriched the class of two-dimensional exactly solvable critical 
models. (sect. 5) I had the great pleasure to be in Leo Kadanoff's group 
in this year. A very short account of Leo's work in the following years,
which is mainly in critical phenomena, is given in section 6.

\section{Critical phenomena}

In the fifties and sixties of the last century a lot of investigations
started to understand the behavior of gas-liquid systems, magnets
and other systems close to their critical point.

The theory by Van der Waals\cite{Waals73} for gas-liquid systems and
that by Curie and Weiss\cite{Weiss07} for magnetic substances
predicted what is called molecular field behavior.
In terms of the paramagnetic-ferromagnetic transition
one describes the critical behavior by
\[
m\propto(-\tau)^{\beta}, \quad \chi\propto |\tau|^{-\gamma},
\quad c\propto |\tau|^{-\alpha}, \quad \xi\propto |\tau|^{-\nu}
\]
at zero magnetic field, where $m$ is the magnetization, $\chi$
the susceptibility, $c$ the specific heat, $\xi$ the correlation
length, and $\tau=(T-T_c)/T_c$ measures the difference of the temperature $T$ to its critical value $T_c$.
For liquid-gas systems $m$ has to be replaced
by the difference $\rho-\rho_c$ of the density and its critical
value. The exponent
$\alpha=0$ corresponds to a jump and/or a logarithmic
divergence of the specific heat.
Molecular field approximation yields a jump in the specific heat and
critical exponents $\beta=1/2$, $\gamma=1$, $\nu=1/2$.

Landau\cite{Landau37} formulated a general theory, which enclosed not only
gas-liquid systems and magnetic materials, but many others
like binary mixtures and the transition to superfluid helium.
He described the systems in terms of an order parameter
which for magnetic systems is the magnetization
and in gas-liquid systems is the difference of the density and the
critical density.

Experimentally one found $\beta\approx 1/3$, $\gamma\approx 4/3$,
and values of $\alpha$ between -0.1 and +0.1.
Part of these observations date back to 1900.
See the review by Levelt-Sengers.\cite{LevSen73}

Two exact solutions showed that the critical behavior might differ
from molecular field behavior:
The three-dimensional spherical\cite{BerKac52} model yields $\gamma=2$ 
and a kink in the specific heat, but no jump, corresponding to
$\alpha=-1$; the two-dimensional
Ising model yields $\beta=1/8$\cite{Yang52} and a logarithmically 
diverging specific heat\cite{Onsager44}.
Moreover, Ginzburg\cite{Ginzburg60} analyzed the fluctuation contributions near the critical point and
came to the conclusion that in three dimensions there is a region
where molecular field theory fails, its temperature-range depending on 
the range of interaction.

A way to obtain the critical exponents was to perform expansions
for lattice models like the Ising model and classical Heisenberg
models in powers of the coupling over temperature. From these expansions
one could estimate the critical temperature and critical exponents.
They came close to experimentally determined exponents. Unfortunately
however, one did not understand the mechanism behind this behavior.

\section{Widom 1965 and Kadanoff 1966}

There were three important steps to understand critical behavior:
The {\it first step} was Ben Widom's homogeneity law\cite{Widom65}
in 1965, according to which
the order parameter is a homogeneous functions of $\tau$ and a second
quantity, which for the gas-liquid transition is the difference
$\mu-\mu_c$ of the chemical potential and its critical value.
For magnets one uses instead the magnetic field $h$.
This was an extremely useful concept. It explained the relations
between critical exponents, which were already known as equalities
or inequalities. Several experiments were 
analyzed accordingly and good agreement was found. Examples are the 
measurements of the magnetization of CrBr$_3$ by Ho and
Litster\cite{HoLit69}
and of nickel by Kouvel and Comly\cite{KouCom68}, and the equation of 
state for various gases by Green, Vicentini-Missoni, and
Levelt-Sengers\cite{GrVMLS67}.

In 1966 Leo Kadanoff\cite{Kadanoff66a} presented the {\it second 
important step} to understand critical
phenomena with his block transformation. He replaced the spins within
a block by a new block spin and introduced an effective interaction
between the block spins, which yields the same behavior for 
magnetization
and spin correlations on distances large in comparison to the block
size. Denote the original coupling and magnetic field by $K$ and $h$
and the block one by $K'$ and $h'$. The number of spins within the block
be $b^d$. Then there is a mapping of $(K,h)\rightarrow(K',h')$:
$K'=f(K)$, and the magnetic field will effectively change by a factor
one may call $b^x$: $h'=b^x h$. It is essential that this
transformation is not
singular, at least not close to the critical coupling $K_c$.
This critical coupling reproduces itself $K_c=f(K_c)$ under the
block transformation. Small deviations from $K_c$, which one
may call $\tau=K-K_c$ increase under the block transformation by a 
factor we may call $b^y$, $\tau'=b^y\tau$.
Then the two exponents $x$ and $y$
determine the critical exponents
\[
\alpha=2-d/y, \quad \beta=(d-x)/y, \quad
\gamma=(2x-d)/y, \quad \nu=1/y.
\]

\section{Interest grows}

In 1967 the interest in critical phenomena had strongly grown.
Several reviews appeared on this subject:
the reviews by
Michael Fisher\cite{Fisher67}, Peter Heller\cite{Heller67},
and Leo Kadanoff\cite{KadGoe67} and his coworkers.

In 1970 I participated in the {\it Midwinter Solid State Research Conference}
on the topic {\it Critical Phenomena}. Among the
participants were many prominent workers in this field. I mention
only G\"unter Ahlers, George Baker, Richard Ferrell, Michael Fisher, 
Robert Griffiths, Bertrand Hal\-pe\-rin, Peter Heller, Pierre Hohenberg, 
Leo Kadanoff, David Landau, J.D. Litster, Paul Martin, Michael 
Schulhoff, Johanna
Levelt-Sengers, Eugene Stanley, Gerard Toulouse, Ken Wilson,
Michael Wortis, and Peter Young.
It was my first travel to the United States and it was a pleasure 
to meet many physicists in person, which before I knew only from their
papers. It was also interesting for me, since I had applied for
a post-doc position with several of them.

Laramore\cite{Laramore70} wrote a short report on this conference and 
concerning static critical phenomena he resumed:
{\it The two-parameter Kadanoff-Widom scaling laws are in real trouble
as far as predicting the relationships between the static critical
exponents in three-dimensional systems, the deviation from the
scaling laws being small, but nevertheless real.}
Thus at that time there were doubts on this theory.
One should be aware that the effect of dipolar interactions and
of anisotropies including crossover effects were not yet sufficiently
clear. However, dipolar effects in uniaxial ferroelectrics were
already considered by Larkin and Khmelnitskii\cite{LarKhm69},
and crossover effects by Jasnow and Wortis\cite{JasWor68}
and by Riedel and myself\cite{RieWeg69}.

\section{Baxter and Wilson 1971}

1971 saw two very important advances in critical phenomena,
the solution of the eight-vertex model by Rodney Baxter and the
renormalization group calculation of Ken Wilson on the basis of
Kadanoff's block spin ideas.

In 1971 I was research-associate in Leo Kadanoff's group.
It turned out that at that time Leo was engaged in urban
planning, but was still interested in critical phenomena.
When I came to Brown University I had already started with
thinking about: How can the idea of duality as introduced
by Kramers and Wannier\cite{KraWan41} for the two-dimensional
Ising model be generalized to higher dimensions?
Kadanoff and Ceva\cite{KadCev71} had
introduced the concept of the disorder variable, which under
duality transforms into the Ising spin. This concept was useful
for my considerations. I realized that the dual model to the
conventional three-dimensional Ising model is an Ising model with 
plaquette interactions. In four dimensions the Ising model with
plaquette interactions is self-dual.\cite{Wegner71}
Thus its transition temperature,
provided there is only one, can be determined in the same way and
with the same result as for the conventional two-dimensional
Ising model. The transition temperature for this model was confirmed
later numerically by Creutz, Jacobs, and Rebbi\cite{CrJaRe79}
to the precision allowed by the hysteresis effects at the first-order
transition. The model has a local gauge invariance.
The products of spins along loops, called Wilson-loops\cite{Wilson74},
show different behavior in the two phases. They decay with increasing
loops with either an area law or a perimeter law, depending on the phase
\[
\langle \prod_{\rm loop} S(r) \rangle \propto 
{\exp(-a/a(T)) \atop \exp(-p/p(T))},
\]
where $a$ is the area, and $p$ the perimeter of the loop.
What I did not realize at that time, was that this behavior is
characteristic for the behavior of confinement and deconfinement
of quarks.

1971 saw the solution of the eight-vertex model by Rodney
Baxter\cite{Baxter71,Baxter72}. The model is equivalent to an
Ising model on a square lattice with two-spin interactions and
a four-spin interaction of strength $\lambda$.
If $\lambda$
vanishes, then the system decays into two conventional Ising models.
Starting from this limit Leo and myself determined the critical exponents
in order $\lambda$.\cite{KadWeg71} This is only possible, since
the system has an operator, which stays marginal, and creates
a line of fixed-points with varying exponents.

Later during that year the papers by Wilson\cite{Wilson71a,Wilson71b}
appeared, in which he derived an approximate renormalization group
equation, which allowed him to determine critical exponents for the
three-dimensional Ising model. This was the {\it third important step}
to understand critical phenomena. These papers were based on
Kadanoff's block spin picture.

Wilson stated that the space of
Hamiltonians is much larger than the one used by Kadanoff, but
Kadanoff has used all relevant operators necessary to reach
the critical point and has thus obtained the correct picture
in the immediate vicinity of the critical point.
Doubts on the Widom-Kadanoff picture soon declined.

Starting from this picture the concept of a fixed point and of
universality classes became clear. Thus all systems whose interactions
converge at criticality to the same fixed point, are governed by
the same critical exponents and -- as later became clear -- also to
the same amplitude ratios. I investigated aspects of the general
structure of the flow of the interaction under renormalization,
considering also irrelevant operators and that the renormalization
includes non-linear flows of the couplings\cite{Wegner72a}.
What assured me that Wilson's approach is the correct approach
was the paper by Wilson and Fisher\cite{WilFish72} on the
$\epsilon$-expansion in $d=4-\epsilon$ dimensions. A theory, which
in complete agreement with the Ginzburg\cite{Ginzburg60} criterion
gave molecular field behavior for $d>4$ and exponents different from
molecular field behavior for $d<4$ has the expected properties.
Wilson's approximate recursion relation was so good that it
gave the correct critical exponents in order $\epsilon$.
We (Grover, Kadanoff and FJW) used it to determine the exponents
for the three-dimensional classical Heisenberg model\cite{GrKaWe72}.
I added a calculation of the critical exponents for the isotropic
$n$-vector model\cite{Wegner72b}.
It happened several times that I did not understand an answer
by Leo immediately, but only one or two days later. But his
answers were so concise that I remembered them after such
a long time.

It is surprising that Ken Wilson developed first this approximate
renormalization group equation and only later came back to the
Landau free-energy functional. The free theory contains only
a two-spin interaction
$\sum_q u(q) S_q S_{-q}$ with (i) $u(q)\propto q^2$ for small
wave-vectors $q$,
and (ii) $u(q)\rightarrow 1$ for $q$ large in comparison to the
cut-off momentum $\Lambda$. The first condition is obvious and can
be easily incorporated in the transformation, but the
formulation of the transformation for the second condition is less
obvious. It can be found in refs. \cite{WilKog74} and \cite{Wegner76a}.

I became also known to Anthony Houghton at Brown University.
A different way which Anthony Houghton, and myself\cite{WegHou73} used, 
was a sharp momentum cut-off.
This procedure works well in first order in $\epsilon$, but yields
potentials non-analytic in the wave-vector for higher orders in
$\epsilon$. We also obtained the behavior of the n-component vector model in the
limit $n\rightarrow\infty$. It agreed with the spherical
model\cite{BerKac52} in accordance with Stanley\cite{Stanley68}.
Our paper became known
as functional renormalization, since we calculated the interaction
as a function.

\section{Later Years}

After my stay at Brown I met Leo only occasionally at conferences.
Once he visited Heidelberg for a talk.

Obviously Leo wanted to learn all secrets of two-dimensional and if
possible three-dimensional critical systems. This includes
correlation functions\cite{Kadanoff66b,Kadanoff69b} as well as operator
algebras\cite{Kadanoff69a,Kadanoff76a} and renormalization group 
procedures in real space,
which are characterized by potential moving and variational
schemes\cite{Kadanoff75,KadHou75,KaHoYa76,ShKaPr79,EfWaKoKa14}.
He studied extensively the renormalization scheme\cite{Kadanoff76b}
by Migdal\cite{Migdal75a,Migdal75b}.
With Jorge Jose, Scott Kirkpatrick, and David Nelson he investigated
the effect of vortices and symmetry-breaking anisotropic fields
on the planar model at low
temperatures\cite{JoKaKiNe77,Kadanoff79}.

But he studied also other systems showing critical behavior as
turbulence\cite{CGHKLTWZZ89}, avalanches\cite{KaNaWuZh89,ChFeKaKoPr93}, 
and the critical behavior of Kolmogorov-Arnold-Moser surfaces
\cite{KadShe82,Kadanoff81,Kadanoff82}, and
mappings, which create critical boundaries.

Some work of Leo\cite{HaJeKaPrSh86,JeKaPr87} together with Halsey, Jensen, Procaccia, and Shrai\-man on fractal 
measures came close to some work of our work on the Anderson transition
in a random potential. At the mobility edge, which separates
extended and localized eigenstates, the states approach a fractal behavior. 
In a block picture one diagonalizes first the interaction
within small blocks. This is continued
with larger and larger blocks. This yields a transformation in real
space and energy space.\cite{Wegner76b}. After the mapping of
this Anderson model to the nonlinear sigma-model was found
\cite{Wegner79,SchWeg80} it was possible to determine
the fractal exponents of the eigenfunctions in $2+\epsilon$ dimensions\cite{Wegner80,HofWeg86,Wegner87}.

Leo characterized his work at several
occasions\cite{Kadanoff90,Kadanoff93,Kadanoff02}.

Unfortunately I missed Leo, when he was in Chicago.
I spent a sabattical at the James-Franck Institute in the summer
of 1978, whereas
Leo moved from Brown University to the James-Frank Institute
in the winter of 1978.
I am glad that I met him
in 2015 at the March meeting of the American Physical Society in
San Antonio. He was still intensively interested in the progress of Physics.

\end{document}